\documentclass[12pt]{article}
\usepackage{amsmath}
\input{epsf}
\def\ltap{\raisebox{-.4ex}{\rlap{$\,\sim\,$}} \raisebox{.4ex}{$\,<\,$}}
\def\gtap{\raisebox{-.4ex}{\rlap{$\,\sim\,$}} \raisebox{.4ex}{$\,>\,$}}

\newcommand\as{\alpha_{\mathrm{S}}}

\def\beq{\begin{equation}}
\def\eeq{\end{equation}}
\def\beeq{\begin{eqnarray}}
\def\eeeq{\end{eqnarray}}

\def\to{\rightarrow}

\begin{document}
\begin{titlepage}
\renewcommand{\thefootnote}{\fnsymbol{footnote}}
\begin{flushright}
    CERN--TH/2001-147   \\ hep-ph/0106049
     \end{flushright}
\par \vspace{10mm}

\begin{center}
{\Large \bf
Higgs production at hadron colliders\\ in (almost) NNLO QCD
\footnote{Talk given by M. Grazzini at the XXXVIth Rencontres de Moriond, QCD and Hadronic interactions, Les Arc1800, France and at the 9th International Workshop on Deep Inelastic Scattering DIS2001, Bologna, Italy.}
\\[1.ex]
}

\end{center}
\par \vspace{2mm}
\begin{center}
{\bf Stefano Catani${}^{(a)}$~\footnote{On leave of absence 
from INFN, Sezione di Firenze, Florence, Italy.}, Daniel de Florian${}^{(b)}$
\footnote{Partially supported by Fundaci\'on Antorchas.}
}
\hskip .2cm
and
\hskip .2cm
{\bf Massimiliano Grazzini${}^{(c,d)}$}\\

\vspace{5mm}

${}^{(a)}$Theory Division, CERN, CH-1211 Geneva 23, Switzerland \\

${}^{(b)}$Institute for Theoretical Physics, ETH-H\"onggerberg, 
CH-8093 Z\"urich, Switzerland \\

${}^{(c)}$ Dipartimento di Fisica, Universit\`a di Firenze, I-50125 Florence, Italy\\

${}^{(d)}$INFN, Sezione di Firenze, I-50125 Florence, Italy

\vspace{5mm}

\end{center}

\par \vspace{2mm}
\begin{center} {\large \bf Abstract} \end{center}
\begin{quote}
\pretolerance 10000

We compute the soft and virtual NNLO QCD corrections to Higgs production
through gluon--gluon fusion
at hadron colliders. We present  numerical results obtained at the LHC and at 
the Tevatron Run II.

\end{quote}

\vspace*{\fill}
\begin{flushleft}
     CERN--TH/2001-147 \\ June 2001 

\end{flushleft}
\end{titlepage}

\setcounter{footnote}{1}
\renewcommand{\thefootnote}{\fnsymbol{footnote}}

%~\vspace{1.cm}

Although the Higgs boson is a crucial ingredient of the Standard Model (SM),
it has so far eluded experimental discovery. LEP results~\cite{LEPres} imply 
a lower limit of $M_H > 113.5$~GeV (at $95\%$ CL) on the mass $M_H$ of the SM
Higgs boson, favour a light Higgs boson ($M_H \ltap 200$~GeV) and suggest the
observation of signal events at $M_H\simeq 115$~GeV.

After the end of the LEP program, the Higgs search will be carried out at hadron 
colliders, where the dominant production mechanism is gluon--gluon fusion through a 
heavy-quark loop.
At the LHC~\cite{LHCest}, $gg$ fusion overwhelms the other production
channels in the case of a light Higgs boson, and it still provides
$\sim 50\%$ of the total production rate at $M_H\simeq 1$~TeV.
At the Tevatron~\cite{Tevest}, $gg$ fusion gives $\sim 65\%$ of the total cross section in the 
mass range $M_H=100$--200~GeV, although this production mechanism leads to an
important discovery mode only at $M_H\gtap 135$~GeV, when the decay channel
$H \to WW^*$ opens up (if $M_H\ltap 135$ GeV, the decay channel 
$H \to b{\bar b}$ suffers from a huge QCD background, and the decay
rate $H\to\gamma\gamma$ is too low to be observed).

%%====================================
\begin{figure}[htb]
\begin{center}
\begin{tabular}{c}
\epsfxsize=9truecm
\epsffile{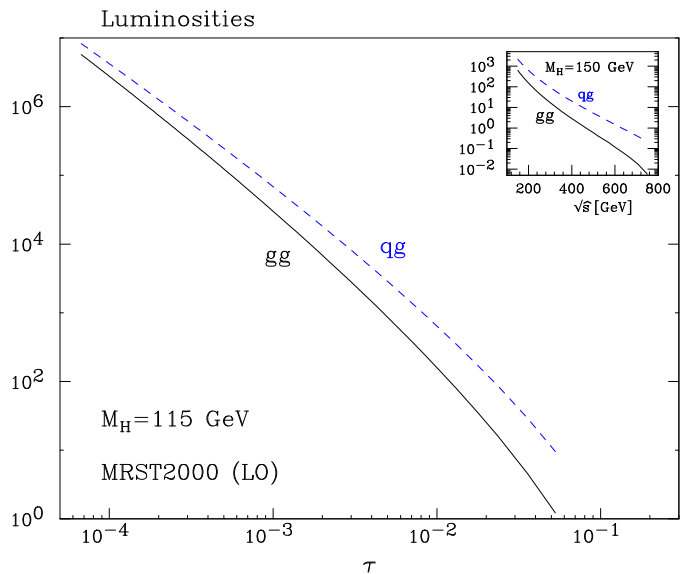}\\
\end{tabular}
\end{center}
\caption{\label{lumi}{Parton luminosities as a function of $\tau={\hat s}/s$ 
at the LHC. The inset plot shows the luminosities at the Tevatron as a function
of the partonic centre-of-mass energy $\sqrt{{\hat s}}$. }}
\end{figure}
%%===================================

The NLO QCD corrections to Higgs boson production through $gg$ fusion
give a large effect~\cite{ggnlo}.
Since approximate evaluations~\cite{Kramer:1998iq} of higher-order terms 
suggest that their effect can still be sizeable, the computation of the NNLO 
corrections is certainly important.

A first step in this direction has been performed by two 
groups~\cite{Catani:2001ic,Harlander:2001is}, who have computed the soft and 
virtual contributions to the NNLO partonic cross section 
${\hat \sigma}(gg\to H+X)$ in the large-$m_{\rm top}$ approximation.
Our calculation~\cite{Catani:2001ic} was done by combining the recent
result~\cite{Harlander:2000mg} for the two-loop amplitude $gg\to H$
with the soft factorization formulae for tree-level~\cite{softtree} 
($gg \to Hgg, Hq{\bar q}$) and one-loop~\cite{softloop} ($gg \to Hg$) amplitudes.
The independent calculation of Ref.~\cite{Harlander:2001is} uses a different
method, and the results for the partonic cross section fully agree.

In Ref.~\cite{Catani:2001ic}, we have combined the NNLO partonic cross section
with the recent MRST2000 set of parton distributions~\cite{Martin:2000gq}, 
which includes (approximate) NNLO densities, thus providing a
first consistent estimate of the NNLO QCD corrections to the hadronic cross
section. In the following, we briefly summarize our main results.
\newpage
%%====================================
\begin{figure}[htb]
\begin{center}
\begin{tabular}{c}
\epsfxsize=12truecm
\epsffile{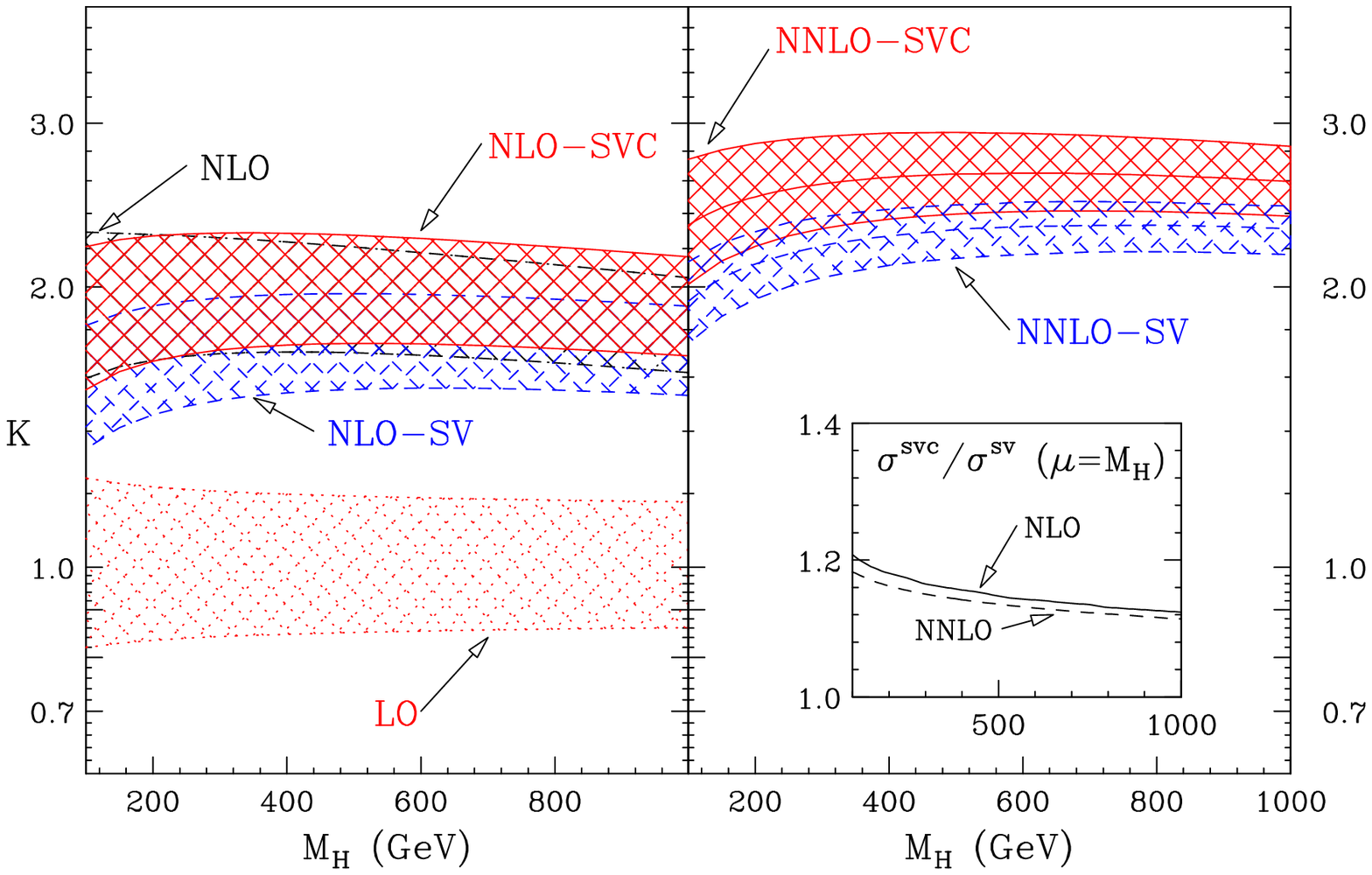}\\
%\end{tabular}
%\end{center}
%\caption{\label{fig:k1}{K-factors at the LHC: exact NLO result, NLO-SV and
%NLO-SVC approximations (left), 
%and NNLO-SV and NNLO-SVC approximations (right).}}
%\end{figure}
%%===================================
%%====================================
%\begin{figure}[htb]
%\begin{center}
%\begin{tabular}{c}
\epsfxsize=12.2truecm
\epsffile{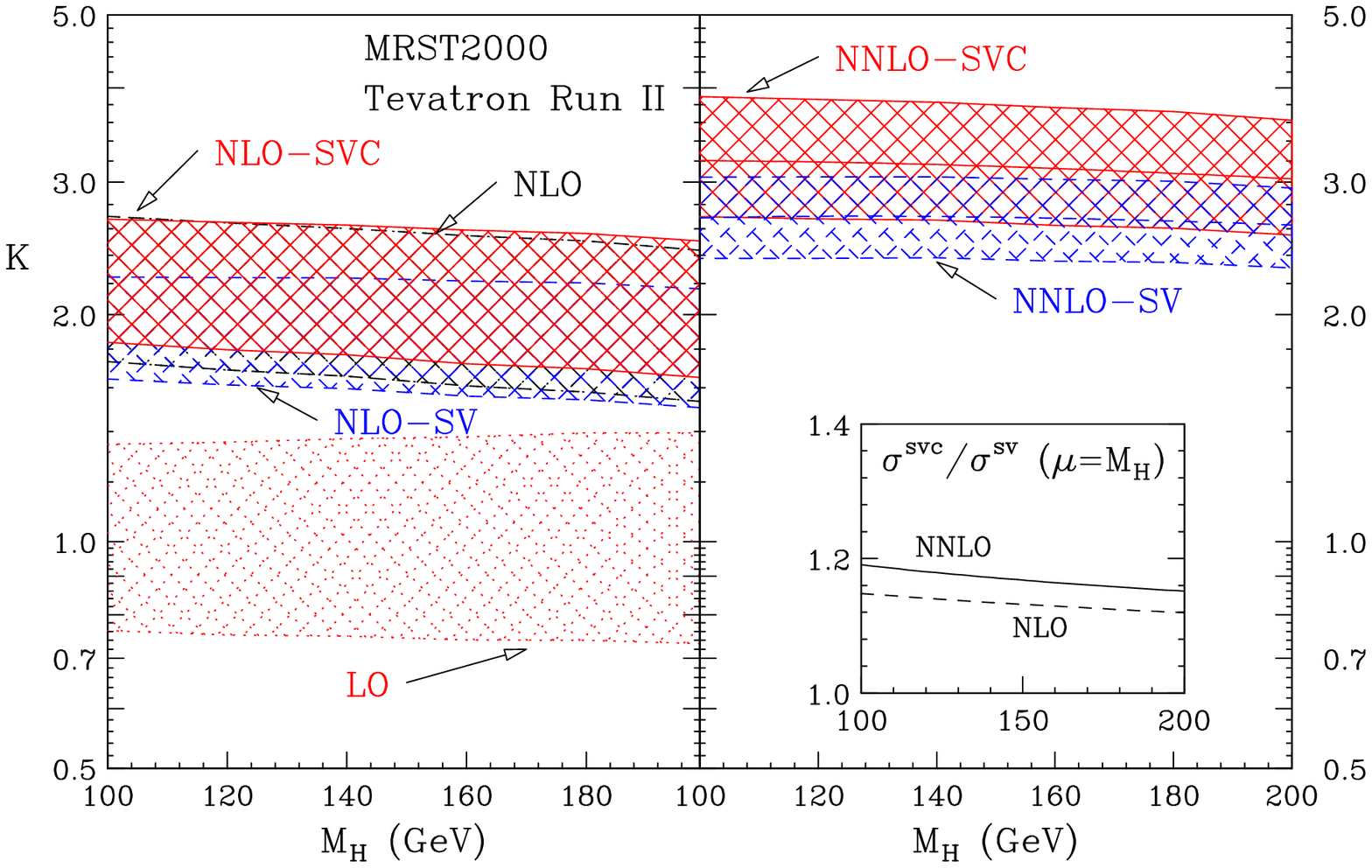}\\
\end{tabular}
\end{center}
%\caption{\label{fig:k2}{K-factors at Tevatron Run II (notation as in 
%Fig.~\ref{fig:k1}).}}
\caption{\label{fig:k1}{K-factors at the LHC (above) and Tevatron Run II 
(below): exact NLO result, NLO-SV and
NLO-SVC approximations (left), 
and NNLO-SV and NNLO-SVC approximations (right).}}
\end{figure}
%%===================================

The soft and virtual terms give the most important contributions to the 
partonic cross section when it is computed at a value of the
partonic centre-of-mass energy
$\sqrt {\hat s}$ that is close to the Higgs boson mass $M_H$. The hadronic 
centre-of-mass energy $\sqrt s$ is related to the partonic one as 
${\hat s}= \tau s = x_1 x_2 s$, where $x_1$ and $x_2$ are the momentum fractions
of the two partons that initiate the partonic subprocess. The density of these
partons is controlled by the parton distributions of the colliding hadrons.
Since the parton distributions are strongly suppressed at large values
of the momentum fractions $x_1, x_2$, ${\hat s}$ is typically much smaller
than $s$ and, therefore, the soft and virtual terms can approximate
the complete result also when the hadronic energy $\sqrt s$ is not very close 
to $M_H$. At fixed $M_H$, the quantitative reliability of the approximation 
depends on the value of $\sqrt s$ and on the actual value of the parton
luminosity. The $gg$ and $(q+{\bar q})g$ luminosities are shown in 
Fig.~\ref{lumi}. At the LHC, where we use the reference value $M_H=115$~GeV,
the parton luminosity decreases by about two orders of magnitude when 
$\sqrt {{\hat s}}$ increases from $M_H$ to $3M_H$ (i.e. when $\tau$ increases 
from $10^{-4}$ to $10^{-3}$). At the Tevatron (see the inset plot),
the parton luminosity decreases by almost two orders of magnitude when 
$\sqrt{{\hat s}}$ increases from $\sqrt{{\hat s}}=M_H=150$~GeV to $300$~GeV.
This observation suggests that the soft and virtual terms can give a good
approximation at the LHC and a (slightly) better approximation at the 
Tevatron Run II.

In Ref.~\cite{Catani:2001ic}, we introduced two approximations, named 
soft--virtual (SV) and soft--virtual--collinear (SVC), of the partonic cross
section. In the SV approximation only the contributions of soft
and virtual origin are taken into account. The SVC approximation extends
the SV approximation by including the next-to-dominant contribution
(which has a collinear origin~\cite{Kramer:1998iq})
in the expansion around the region where $\sqrt{{\hat s}}=M_H$.

In the following we present numerical results for the hadronic cross section
evaluated up to LO, NLO and NNLO. At the NLO we compare the exact result
with those of the SV and SVC approximations. At the NNLO, we add the exact 
NLO result to our SV and SVC approximations of the NNLO corrections.
All the results are presented in term of K-factors, 
which are defined as the
ratio of the hadronic cross section over its LO value. The latter
is computed by fixing the factorization and renormalization scales
$\mu_F$ and $\mu_R$ at the central value $\mu_F=\mu_R=M_H$.
As a reference, we use the parton distributions of the MRST2000 set,
with parton densities and $\as$ evaluated at each corresponding order. 

At the LHC, we find that the results do not strongly depend on the choice
of parton distributions: very similar results are obtained by using the
CTEQ5 set~\cite{cteq}, whereas larger ($\sim 10\%$) deviations appear by 
using the GRV98 set~\cite{grv}. At the Tevatron Run II,
the dependence on the parton distributions is stronger than at the LHC. 
The CTEQ5 set gives a LO cross section at $M_H=100$~GeV ($M_H=200$~GeV)
that is $\sim 10\%$  ($\sim 30\%$) smaller than the value obtained by using the
MRST2000 set. Of course, this dependence affects the K-factors through their
normalization with respect to the LO cross section.

The K-factors at the LHC and at the Tevatron are shown in
Fig.~\ref{fig:k1}.
% and \ref{fig:k2}, respectively. 
The bands are obtained by independently varying 
$\mu_F$ and $\mu_R$ in the range $M_H/2 \le \mu_F, \mu_R \le 2M_H$.
The results show a reduction of the scale dependence at NNLO 
(from about $\pm 20 \%$ at full NLO to about $\pm 10 \%$ and $\pm 15 \%$
at NNLO-SV and NNLO-SVC, respectively).

From the plots on the left-hand side of Fig.~\ref{fig:k1},
% and \ref{fig:k2}, 
we see that at NLO the SV approximation tends to underestimate the exact result,
whereas the SVC approximation slightly overestimates it. Therefore, 
it is reasonable to expect the exact NNLO K-factor to lie inside the 
corresponding SV and SVC bands. In the case of the production of a light Higgs 
boson ($M_H=100$--200~GeV)
%($100~{\rm GeV} \ltap M_H \ltap 200~{\rm GeV}$) 
at the LHC (Tevatron),
this expectation corresponds to $K\simeq 2.2$--$2.4$ ($K\simeq 3$), i.e. to
an enhancement of the NLO cross section by about 10--25$\%$ ($50\%$).

Note that, at fixed $M_H$, Higgs boson production at the Tevatron
is closer to threshold than at the LHC. This implies 
(see Fig.~\ref{fig:k1}) 
%and \ref{fig:k2}) 
that the approximation
of the fixed-order results in terms of the corresponding SV and SVC
contributions works better at the Tevatron than at the LHC. This also implies
that, order by order in perturbation theory, QCD radiative corrections are
larger at the Tevatron than at the LHC. In particular, QCD contributions
beyond NNLO, due to multiple soft-gluon emission, can still be numerically
relevant to Higgs boson production at the Tevatron.

\noindent {\bf Acknowledgments.}\\
\noindent This work was supported in part 
by the EU Fourth Framework Programme ``Training and Mobility of Researchers'', 
Network ``Quantum Chromodynamics and the Deep Structure of
Elementary Particles'', contract FMRX--CT98--0194 (DG 12 -- MIHT).

\end{document}